\documentclass[journal,lettersize]{IEEEtran}
\usepackage[usenames]{color}
\usepackage{amsfonts}
\usepackage{cite}
\usepackage{amssymb,amsthm}
\usepackage{bbm, amsmath}
\usepackage[caption=false]{subfig}
\usepackage{graphicx}
\usepackage{footnote}
\usepackage{algpseudocode}
\usepackage{algorithm}
\usepackage{multirow}
\usepackage{amsmath}
\usepackage{xcolor}
\usepackage{verbatim}
\usepackage{times}
\usepackage{epstopdf}
\usepackage{bm}
\usepackage{color,soul}
\usepackage{tikz}
\usepackage{pgfplots}
\pgfplotsset{compat=1.17}
\usetikzlibrary{shapes,arrows,positioning,fit,backgrounds,matrix,calc,trees,arrows.meta}
\usetikzlibrary{patterns}
\usepackage{balance}
\usepackage{pifont}
\usepackage{booktabs}

\allowdisplaybreaks[4] 

\title{Toward a Sustainable Federated Learning Ecosystem: A Practical Least Core Mechanism for Payoff Allocation}

\author{Zhengwei~Ni,~\IEEEmembership{Member,~IEEE,} Zhidu~Li,~\IEEEmembership{Member,~IEEE,} Wei~Chen,~\IEEEmembership{Member,~IEEE,}
Zhaoyang~Zhang,~\IEEEmembership{Senior Member,~IEEE,} Zehua Wang,~\IEEEmembership{Member,~IEEE,} F.~Richard~Yu,~\IEEEmembership{Fellow,~IEEE,} and Victor~C.~M.~Leung,~\IEEEmembership{Life Fellow,~IEEE}
	\thanks{The work of Zhengwei Ni was supported in part by the Tongxiang General Artificial Intelligence Research Institute Project TAGI2-B-2024-0017 and in part by Zhejiang Provincial Natural Science Foundation of China under Grant No.LMS26F010017. The work of Zhidu Li was supported in part by Science and Technology Research Program of Chongqing Municipal Education Commission, grant number KJQN202300646. The work of Wei Chen was supported in part by by the National Natural Science Foundation of China under Grants 52274160 and 51874300,  ``Jiangsu Distinguished Professor'' Project in Jiangsu Province (140923070).}
		
	\thanks{Zhengwei Ni is with the School of Information and Electronic Engineering (Sussex Artificial Intelligence Institute), Zhejiang Gongshang University, Hangzhou 310018, China, and also with the School of Computer Science and Technology, University of Mining and Technology, Xuzhou 221116, China (emails: zhengwei.ni@zjgsu.edu.cn).}
	\thanks{Zhidu Li is with School of Communications and Information Engineering, Chongqing University of Posts and Telecommunications, Chongqing 400065, China (emails: izd@cqupt.edu.cn).}
	\thanks{Wei Chen is with the School of Computer Science and Technology, University of Mining and Technology, Xuzhou 221116, China, and also with the School of Mechanical, Electrical and Information Engineering, China University of Mining and Technology (Beijing) and Key Laboratory of Intelligent Mining and Robotics, Ministry of Emergency Management, Beijing 100083, China (emails: chenwdavior@163.com).}
	\thanks{Zhaoyang Zhang is with the College of Information Science and Electronic Engineering, Zhejiang University, Hangzhou 310027, China; the Key Laboratory of Information Processing, Communication and Networking of Zhejiang Province (IPCAN), Hangzhou 310027, China; the Key Laboratory of Collaborative Sensing and Autonomous Unmanned Systems of Zhejiang Province, Hangzhou 310015, China; and the International Joint Innovation Center, Zhejiang University, Haining 314400, China (email: zhzy@zju.edu.cn).}
	\thanks{Zehua Wang is with the Department of Electrical and Computer Engineering, The University of British Columbia, Vancouver, BC V6T 1Z4, Canada(e-mail: zwang@ece.ubc.ca).}
	\thanks{F. Richard Yu is with Systems and Computer Engineering, Carleton University, Ottawa, ON, Canada. (e-mail: Richard.Yu@carleton.ca).}
	\thanks{Victor C. M. Leung is with the College of Computer Science and Software Engineering, Shenzhen University, China, and also with the Department of Electrical and Computer Engineering, the University of British Columbia, Vancouver, Canada (e-mail: vleung@ieee.org).}
	\thanks{Wei Chen is the corresponding author.}

}

\IEEEoverridecommandlockouts

\begin{document}
\maketitle
\newtheorem{mydef1}{Lemma}
\newtheorem{mydef2}{Theorem}
\newtheorem{mydef3}{Remark}
\newtheorem{mydef4}{Definition}
\newtheorem{mydef5}{Proposition}
\vspace{-15mm}
\begin{abstract}
Emerging network paradigms and applications increasingly rely on federated learning (FL) to enable collaborative intelligence while preserving privacy. However, the sustainability of such collaborative environments hinges on a fair and stable payoff allocation mechanism. Focusing on coalition stability, this paper introduces a payoff allocation framework based on the least core (LC) concept. Unlike traditional methods, the LC prioritizes the cohesion of the federation by minimizing the maximum dissatisfaction among all potential subgroups, ensuring that no participant has an incentive to break away. To adapt this game-theoretic concept to practical, large-scale networks, we propose a streamlined implementation with a stack-based pruning algorithm, effectively balancing computational efficiency with allocation precision. Case studies in federated intrusion detection demonstrate that our mechanism correctly identifies pivotal contributors and strategic alliances. The results confirm that the practical LC framework promotes stable collaboration and fosters a sustainable FL ecosystem.
\end{abstract}

{}

\section{Introduction} \label{intro}
By integrating advanced capabilities in communications, computation, and artificial intelligence (AI), next-generation networks such as 6G are expected to enable a wide range of intelligent applications and services—from smart cities to highly automated network management and orchestration. In this landscape, federated learning (FL) has emerged as a promising approach for collaboratively training machine learning (ML) models without sharing raw data, thereby preserving data privacy. 

Especially, the practical applications of cross-silo FL, where the participants are typically companies or organizations, are rapidly expanding \cite{Cross-Silo_Federated_Learning}. For instance, as shown in \cite{Cross-Silo_IoRT_Environments}, FL can be applied in the cross-silo Internet of Robotic Things (IoRT) to allow large-scale, heterogeneous robots to jointly train intelligent models. This significantly enhances production efficiency, operational quality, and collective decision-making in complex industrial environments. Another example is the use of FL in network slicing, which supports collaborative attack detection while maintaining data privacy and essential isolation between different network slices \cite{6G-V2X_Network_Slicing}. However, the successful construction of ML models relies heavily on the quality and quantity of data provided by its participants—be they mobile operators, enterprises, or other entities. Therefore, to build a sustainable FL ecosystem, a central challenge is designing a payoff allocation mechanism that is both efficient and stable.

To address this challenge, several payoff allocation mechanisms have been explored in the literature \cite{Towards_Fairness-Aware_Federated_Learning}. Simpler methods, such as those based on self-reporting \cite{self_reporting} or reputation \cite{Incentive_Mechanism_for_Reliable_Federated_Learning}, are straightforward to implement. However, their effectiveness hinges on the assumption of trustworthy participants and the existence of a credible evaluation metric. Other methods, such as marginal loss or leave-one-out \cite{ghorbani2019data}, determine payoffs by estimating the profit loss when a single participant departs from the coalition—typically the grand coalition. However, these approaches narrowly focus on individual contributions within a specific coalition structure. The Shapley value (SV) improves upon these by averaging each participant’s marginal contribution across all possible coalition formations, yet it suffers from high computational complexity and may require other approximation techniques, which limits its scalability in large-scale settings \cite{WTDP-Shapley,Greedy_Shapley}.

To address these limitations, in this paper, we apply the idea of the least core (LC) to FL as an innovative approach for payoff allocation. By relaxing the stringent requirements of the core solution while allowing for a bounded level of dissatisfaction, we propose a practical and scalable payoff allocation mechanism. By aiming to satisfy all participants as equitably as possible, our proposed LC-based mechanism promotes stable collaboration and fosters a sustainable FL ecosystem.

\section{Least Core in Federated Learning Network}\label{section:II}

The promise of FL in applications like collaborative intrusion detection hinges on a robust incentive model. This model is the economic engine that drives participation, ensuring that organizations or companies are fairly compensated for the value they bring. But this immediately begs the question: in a world where not all data is created equally important, how do we distribute the fruits of this collaboration in a way that is both fair and sustainable, ensuring the federation remains intact?

The most intuitive—and most misleading—metric is data size. This common pitfall ignores the stark reality of data valuation: a small, curated dataset highlighting a rare but critical anomaly is exponentially more valuable than gigabytes of routine, redundant data. A framework that rewards volume over insight not only misallocates profits but also discourages the contribution of the very data that makes the collaboration worthwhile.

Other indirect metrics, such as the technical distance of model updates \cite{Privacy-Preserving_Blockchain-Based} or reputation systems \cite{Incentive_Mechanism_for_Reliable_Federated_Learning}, attempt to move beyond this simplistic view. Yet, while computationally convenient, they remain poor proxies for true value. What the industry truly needs is a mechanism that directly links rewards to real-world profit gains, ensuring that payoffs are allocated not by their size or technical minutiae, but by their ultimate impact on the model's overall profit.

Hence, to achieve this goal, the fundamental principle for paying the participants involved in FL is simple: participation must be more rewarding than doing it alone. This means that any single participant, or even a subgroup, must not have an incentive to break away and form a competing coalition. However, in many complex, real-world scenarios, designing such a self-enforcing and stable reward allocation can be mathematically infeasible—a scenario referred to in game theory as an ``empty core''.

However, certain critical applications—such as public safety communications —demand cooperation essential for societal well-being. For example, infrastructure and spectrum sharing between public safety networks and commercial systems can significantly enhance the effectiveness of emergency response networks [17]. In such contexts, all operators should remain committed, and the departure of even one party could jeopardize the shared objective of ensuring public safety. 

This is where our LC approach offers a practical alternative. The LC moves beyond the search for a perfect, unbreakable allocation. Instead, it allows for some level of `dissatisfaction' among participants. By minimizing the dissatisfaction of the most-tempted breakaway group, it finds the ``most stable'' possible solution. It thus provides a more flexible and realistic framework for fairly payoff allocation—especially in large-scale, industrial FL deployments.

The power of the LC framework, however, extends beyond navigating these fragile, ``empty core'' scenarios. In many emerging industrial FL ecosystems, cooperation is inherently synergistic. This means any group of participants collaborating generates more value than the sum of their individual efforts—a classic win-win situation known in game theory as a ``superadditive game''.

In these favorable contexts, the interpretation of the LC elegantly shifts. Instead of minimizing the maximum dissatisfaction for the most tempted-to-leave party, its objective becomes to maximize the minimum satisfaction across all possible coalitions. In essence, the LC focuses on the "cooperative surplus"—the additional value created by teamwork—and allocates it to ensure that even the coalition with the smallest share of this gain is rewarded as generously as possible. This approach builds a foundation of trust and fairness, guaranteeing that every partner, regardless of their size or contribution, receives a meaningful and motivating reward, thus preventing the marginalization of smaller players and cementing the long-term stability of the alliance.

This duality establishes the LC as an exceptionally robust and versatile framework for industrial FL. Whether the value function in a federated learning network exhibits diminishing returns or superadditivity, the LC provides a unified mechanism for designing incentive-compatible payoff allocations. Its adaptability is crucial for fostering the large-scale, heterogeneous federated ecosystems of the future, which must thrive under diverse and dynamic market conditions.

\section{Implementation of Practical LC}
While the theoretical foundation of the Least Core (LC) offers a compelling framework for fair payoff allocation, its direct implementation faces a critical scalability challenge in real-world federated learning (FL) systems. The standard LC approach requires evaluating the payoff—typically tied to model performance—that every possible subset of participants could achieve by training an FL model on their own. This implies training a separate model for every coalition, a process whose computational cost grows exponentially with the number of participants. While manageable for small federations of two or three operators, this becomes entirely impractical in industrial-scale deployments involving 15 to 20 participants, where the number of required models could range from tens of thousands to millions.

\begin{figure}[!th]
	\centering
	\includegraphics[scale=0.45]{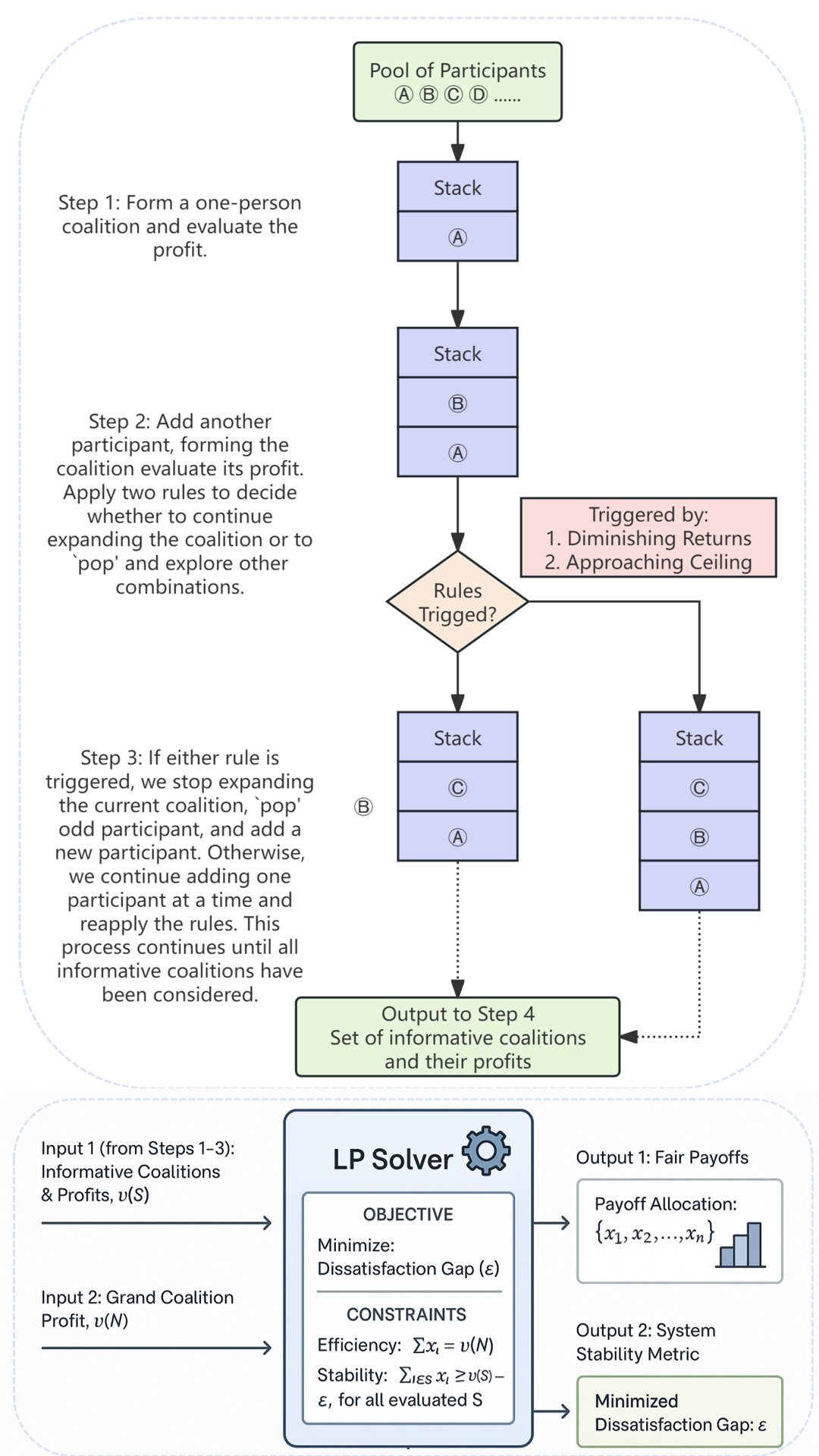}
	\caption{Implementation of the practical LC scheme. \textbf{(Top)} Steps 1-3, which use the LIFO principle to find tall informative coalitions. \textbf{(Bottom)} Step 4, illustrating the structure of our LP solver for payoff allocation.}\label{LC_scheme}
\end{figure}

To bridge this gap between theory and practice, we propose a streamlined deployment scheme that intelligently minimizes the number of coalitions to be evaluated, while ensuring the resulting payoff allocation closely approximates the full LC solution. Our mechanism is inspired by the `stack' data structure in computer science, which operates on a last-in, first-out (LIFO) principle. To provide an intuitive understanding, the entire process is visualized in Figure \ref{LC_scheme}. The procedure is elaborated as follows:

\textbf{Step 1:} We begin by `pushing' a single participant, say A, onto the stack—forming a one-person coalition—and evaluate the profit $v(\{\rm{A}\})$ it can achieve alone.

\textbf{Step 2:} We then add another participant, B, forming the coalition \{A,B\}, and evaluate its profit $v(\{\rm{A,B}\})$. This mirrors real-world business engagements: we first assess individual capabilities, then small partnerships, before scaling to more complex alliances. At this stage, we apply two key rules to decide whether to continue expanding the coalition or to `pop' B and explore other combinations:
\begin{itemize}
	\item \textbf{Rule 1: The Law of Diminishing Returns.} If adding a new participant to an existing coalition provides only a negligible profit increase (i.e., the marginal profit from adding B, $v({\rm \{A,B\})}-v({\{\rm A\})}$, falls below a predefined threshold $t_1$), we can infer that further expanding this coalition is unlikely to yield significant value. This is grounded in FL research, which consistently shows diminishing marginal gains as coalitions grow [11, Fig.3]. By pruning this path, we avoid redundant evaluations. Importantly, any `missed' potential from a participant like C is still captured when C is evaluated individually or in other small coalitions (e.g., \{A,C\} or \{B,C\}), ensuring no critical contributor is overlooked.
	\item \textbf{Rule 2: Approaching the Performance Ceiling.} If a coalition already achieves a profit level that is exceptionally close to the total profit of the entire federation (i.e., the difference between $v(\rm{\{A,B\}})$ and the full federation's profit is less than the predefined threshold $t_2$), then evaluating its larger supersets becomes unnecessary—they cannot meaningfully outperform what has already been achieved. This path is also pruned.
\end{itemize}

\textbf{Step 3:} If either rule is triggered, we stop expanding the current coalition, `pop' participant B, and add a new participant C to form \{A,C\}. Otherwise, we continue adding one participant at a time (i.e., forming \{A,B,C\}) and reapply the rules. This process continues until all informative coalitions have been considered.

\textbf{Step 4:} After our algorithm has efficiently identified the profits of the most critical coalitions, the final step is to translate this information into a fair and stable payoff allocation. We achieve this by formulating a linear program (LP)—a standard and highly efficient optimization tool. The LP's objective is to distribute the total profit generated by the full federation (grand coalition) in a way that minimizes the dissatisfaction of the most-tempted-to-leave subgroup among those we evaluated.

The LP solver is tasked with determining two sets of values:
\begin{enumerate}
	\item \textbf{The Payoff Allocation:} A specific profit share for each participant. These are the primary outputs of our framework.
	\item \textbf{The Minimized Dissatisfaction Gap:} A single value representing the smallest possible dissatisfaction level that any subgroup must accept. This can be conceptualized as a ``fairness tax''—an unavoidable cost of maintaining the grand coalition, which our model makes as low as possible for everyone. In the superadditive (win-win) scenarios described earlier, this `tax' can even become a `bonus'. Then, this value can be negative, representing a guaranteed minimum share of the cooperative surplus that every coalition receives, powerfully incentivizing participation.
\end{enumerate}

The LP solver operates under two simple, yet powerful, constraints:

\begin{itemize}
	\item \textbf{Efficiency:} The sum of all individual payoffs must equal the total profit of the full federation—ensuring all value is distributed. It is called efficiency in game theory.
	\item \textbf{Stability:} For every coalition we evaluated, its members' combined payoff must be at least as good as what they could achieve on their own, minus the minimal, system-wide ``fairness tax''. This constraint is the heart of the LC logic. It mathematically guarantees that no subgroup feels significantly disadvantaged by remaining in the federation, thus cementing the alliance's stability.
\end{itemize}

By following these steps, we transform the payoff allocation into a concrete and solvable engineering problem. This practical approach makes the LC not just a compelling theoretical idea, but a powerful, scalable tool for building the stable, efficient, and sustainable FL ecosystems of the future.\begin{figure*}[!th]
	\centering
	\includegraphics[scale=0.15]{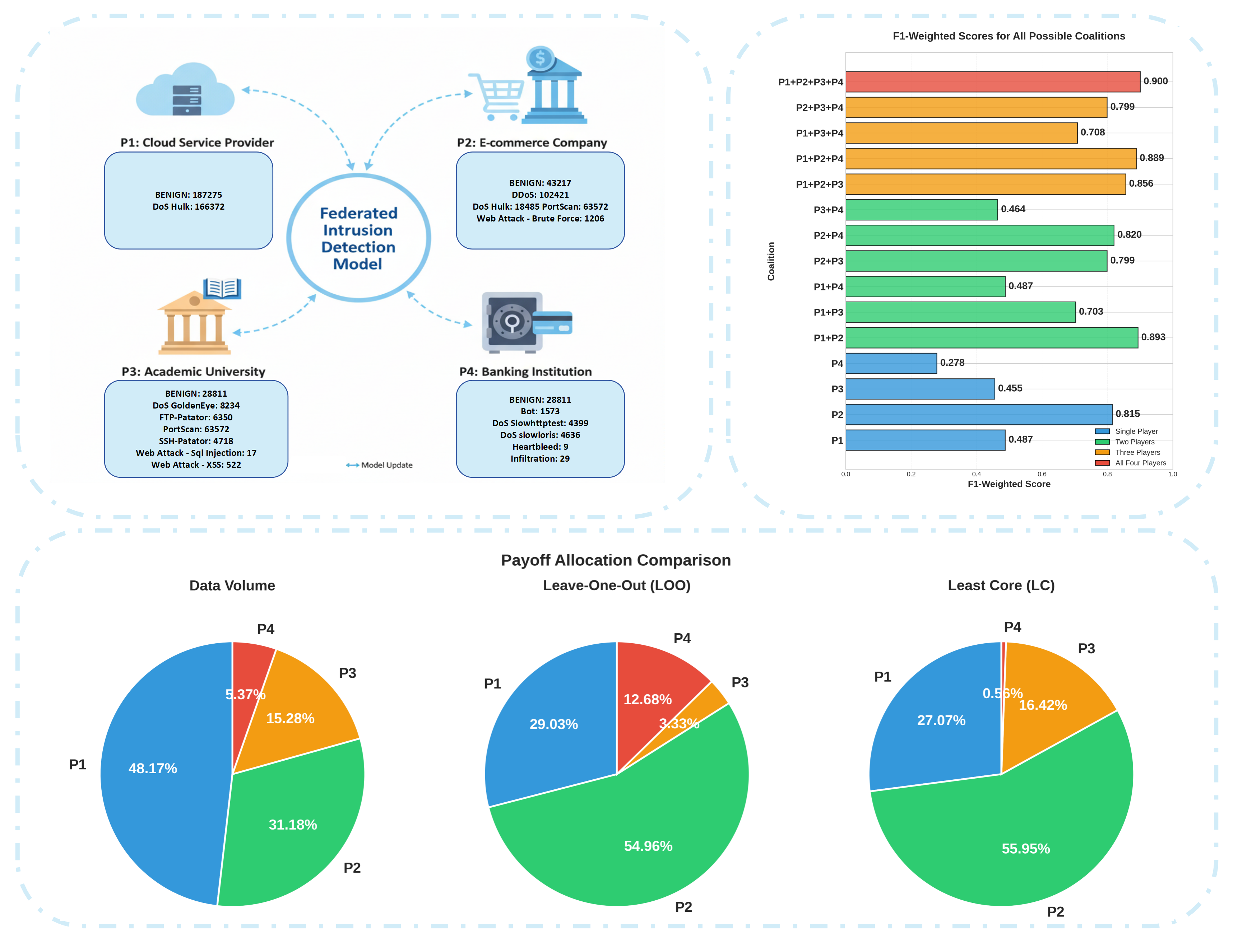}
	\caption{Overview of the federated intrusion detection case study and payoff allocation results. \textbf{(Top-left)} The four participants with their distinct organizational profiles and data distributions. \textbf{(Top-right)} The F1-weighted performance scores for all 15 possible coalitions, which define the value of each partnership. \textbf{(Bottom)} A comparative analysis of final payoff allocations under three methods: allocation by data volume, the leave-one-out, and our proposed least core method.}\label{fig2}
\end{figure*}

\section{Commercial Case Study: Federated Intrusion Detection Model}

To validate our proposed framework, we consider a realistic commercial scenario in which four distinct organizations collaborate to build a federated intrusion detection model using their respective attack detection records. We simulate this setup using a tailored version of the CICIDS-2017 dataset \cite{sharafaldin2018toward}, which reflects real-world network traffic and attack patterns.

The four participants include:
\begin{itemize}
	\item P1: a cloud service provider,
	\item P2: an e-commerce company,
	\item P3: an academic university,
	\item P4: a banking institution.
\end{itemize}

Due to their divergent operational profiles, each organization encounters different types of cyber threats. For example, the academic university (P3) frequently faces \textit{FTP-Patator} attacks, as many universities still maintain FTP servers for small-scale data exchanges. In contrast, the e-commerce company (P2) is more susceptible to \textit{Web Attack – Brute Force} attempts, where attackers target login pages to compromise user accounts through credential stuffing and brute-force techniques.

We model the total profit generated by the federated model as proportional to its F1-weighted score—a balanced metric that harmonizes precision and recall, thus reflecting real-world business value more accurately than accuracy alone. The detailed experimental setup is available in Fig. \ref{fig2}.

We now present the final payoff allocation results obtained using our proposed LC method, alongside a comparative analysis with two baseline approaches: the leave-one-out (LOO) method and allocation based purely on data volume. Since there are only 4 participants, in the LC, we evaluate all possible coalitions, with an acceptable computational cost, for a better analysis.

The data volume metric, which allocates nearly half the reward (48.17\%) to P1, proves to be economically irrational. This approach heavily over-rewards participants with large volumes of redundant, low-value data, while severely undervaluing those who contribute rare, high-impact samples. Such a scheme would disincentivize specialized participants from joining the federation, as their critical—albeit smaller—datasets would be marginalized. For instance, P2 (E-commerce), which achieves a high standalone F1-score thanks to its high-quality data, receives a significantly smaller share than P1. This  clearly illustrates that rewarding volume fails to reflect a participant's actual impact on model performance.

The LOO method offers an improvement over the data volume approach by focusing on performance, but its calculation is fundamentally myopic. It exclusively considers a participant's marginal contribution to the grand coalition (i.e., the performance drop when that single participant leaves). From this narrow perspective, P4's departure causes a larger drop than P3's, leading LOO to assign P4 a greater payoff. However, this ignores the broader strategic landscape. P4 possesses limited bargaining power; while its data is unique, it is insufficient to build a potent model on its own ($v(\{\rm{P4}\})=0.2784$) or to add significant value to smaller partnerships. The LC algorithm correctly identifies this weak negotiating position, recognizing that P4 lacks a credible ``outside option'' and cannot realistically threaten to leave.

In contrast, the LC framework's strategic brilliance lies in its ability to identify P3's pivotal role in forming powerful synergistic sub-coalitions. The coalition \{P1, P3\}, for example, achieves a remarkable F1-weighted score of 0.7029—a dramatic improvement over the individual efforts of P1 (0.4874) and P3 (0.4554). This potent combination of P1's vast background data and P3's diverse, non-commercial attack types creates a strong, independent model, representing a credible breakaway threat. If P3 were under-rewarded (as it would be under LOO), the {P1, P3} coalition would be highly incentivized to defect. The LC framework anticipates this instability and proactively allocates a larger share of the profit to P3. This is not merely a reward for P3's data, but a strategic investment to mitigate the most potent breakaway threat, thereby ensuring the stability of the entire federation.

To further demonstrate the scalability and practicality of our proposed LC scheme, we extended the experimental setup to simulate a larger-scale federation. By subdividing each of the original four participants into four equal sub-entities, we constructed a scenario comprising 16 participants. In a traditional full-evaluation framework, this would necessitate training $2^{16}-2 = 65534$ distinct models, which is a computationally prohibitive task. In stark contrast, our pruning algorithm significantly reduces this computational burden. By setting the thresholds to $t_1 = 0.1$ and $t_2 = 0.1$, the number of required model evaluations drops precipitously to just $97$. Relaxing the constraints slightly to $t_1 = 0.15$ and $t_2 = 0.15$ further reduces the number of trained models to $71$. This reduction demonstrates that our method makes payoff allocation feasible even in larger networks. Guidelines for selecting appropriate thresholds based on specific operational contexts are summarized in Table \ref{table1}.

\begin{table*}[!h]
	\centering
	\caption{Guidelines for Selecting Pruning Thresholds $t_1$ and $t_2$ in Practical LC Implementation.}
	\resizebox{0.9\textwidth}{!}{\begin{tabular}{p{2.5cm}p{4cm}p{5cm}}
		\toprule
		\textbf{Thresholds} & \textbf{Scenario} & \textbf{Expected Outcome} \\
		\midrule
		$t_1, t_2 \to 0$ & High-stakes applications, a small number of participants (2-5) & High accuracy, high computational cost. Close to exact LC. \\
		\addlinespace
		Moderate $t_1$, $t_2$ & General-purpose FL networks with moderate number of participants (5-20) & Balanced trade-off between accuracy and efficiency. \\
		\addlinespace
		High $t_1$, $t_2$ & a large number of participants ($>$20) & Low computational cost, good for identifying top contributors, but may not be accurate. \\
		\bottomrule
	\end{tabular}}\label{table1}
\end{table*}

\section{Least Core vs. Shapley Value}

Both the LC and the SV determine payoff allocations by first assessing the value generated by the subsets of participants. Given this common foundation, a critical question arises: how exactly do the allocation mechanisms of these two approaches differ? To illuminate these differences, this section analyzes another commercial scenario—smart healthcare—to reveal potential divergences in their payoff allocations.

In detail, we simulate a collaborative smart healthcare scenario for heart disease prediction. This simulation is built upon the well-known UCI Cleveland Heart Disease Dataset \cite{heart_disease_45} and implemented using a vertical federated learning (VFL) framework. The dataset contains 13 features, ranging from demographic information (e.g., age, sex) to clinical results (e.g., resting blood pressure), making it ideal for modeling a multi-agency diagnostic process. In our simulated ecosystem, the dataset's features are vertically partitioned among three distinct participants, each representing a different entity in the healthcare value chain:

\begin{itemize}
	\item \textbf{Participant a (Demographics Unit)} owns only two features: `age' and `sex'. This represents a department or agency that collects basic patient data without performing any medical examinations.
	\item \textbf{Participant b (General Clinic)} owns six features, such as ``resting blood pressure'' and ``fasting blood sugar''. This simulates a general medical screening center, but not specialized in cardiology.
	\item \textbf{Participant c (Cardiology Specialist)} owns the remaining five features, including ``exercise-induced angina'' and ``the slope of the peak exercise ST segment''. This corresponds to a specialized cardiology department that conducts heart disease-related tests.
\end{itemize}

In this scenario, we set the profit of the trained model proportional to its recall. Recall is a crucial metric in medical diagnostics because it quantifies the proportion of actual positive cases (patients with heart disease) that are correctly identified. A high recall value signifies the model's ability to effectively identify the majority of patients with heart disease, thus minimizing false negatives. False negatives, in this context, represent instances where patients with heart disease are incorrectly classified as healthy, potentially leading to missed diagnoses and delayed treatment. In medical diagnostics, especially for life-threatening conditions like heart disease, false negatives are significantly more detrimental than false positives (healthy individuals misclassified as patients) due to the risk of delaying critical care.

Here we employ a synchronous vertical FL algorithm, FedBCD \cite{FedBCD}, and conduct the training with all individual participants and their possible combinations. A representative experimental run yielded the following recall values, which serve as the profit for our analysis: $v\left(\{{\rm a}\}\right)=0.5$, $v\left(\{{\rm b}\}\right)=0.6071$, $v\left(\{{\rm c}\}\right)=0.8214$, $v\left(\{{\rm a,b}\}\right)=0.6429$, $v\left(\{{\rm a,c}\}\right)=0.7857$, $v\left(\{{\rm b,c}\}\right)=0.8214$, $v\left(\{{\rm a,b,c}\}\right)=0.8571$. These values form the basis for calculating the payoff allocation results, using the LC and the SV respectively. The final allocation results are presented as Table \ref{Table_VFL}.

\begin{table}[!h]
	\centering
	\caption{Contribution evaluation by the LC and the SV.}\label{Table_VFL}
	\resizebox{0.4\textwidth}{!}{\begin{tabular}{|c|c|c|c|}
		\hline
		{} & ${\rm a}$ & ${\rm b}$ & ${\rm c}$ \\
		\hline
		Least Core & 0.1429 & 0.25 & 0.4643 \\
		\hline
		Shapley Value & 0.1786 & 0.25 & 0.4285 \\
		\hline
	\end{tabular}}
\end{table}

\begin{figure}[!h]
	\centering
	\hspace{-8.7mm}
	\includegraphics[scale=0.65]{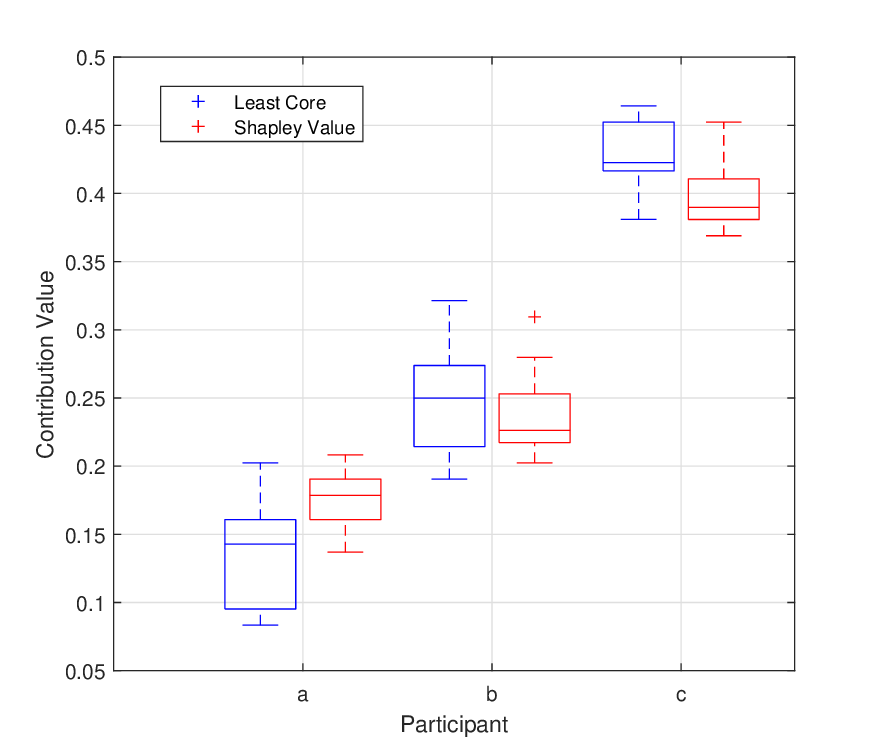}
	\caption{The contribution of each participant evaluated by the LC and the SV.}\label{LV_SC}
\end{figure}

From Table \ref{Table_VFL}, we can find that a divergence emerges in the payoff allocation provided by the LC and the SV. Specifically, the LC assigns a comparatively higher contribution to Participant ${\rm c}$ (0.4643) and a lower contribution to Participant ${\rm a}$ (0.1429) than the Shapley Value (${\rm c}$: 0.4285, ${\rm a}$: 0.1786). This discrepancy shows the difference in how these two approaches conceptualize and quantify contribution in FL networks.

LC prioritizes the concept of `indispensability' for achieving a certain performance threshold. While Participant ${\rm a}$ or ${\rm b}$ alone, or even together, yield relatively lower recall scores, the inclusion of Participant ${\rm c}$ significantly improves the performance. This indicates that Participant ${\rm c}$ possesses features that are crucial for reaching a high recall level. By aiming to minimize the deficit defined before, LC should ensure that no coalition feels significantly under-compensated, inherently emphasizes the participants whose absence severely degrades the overall outcome. Therefore, Participant ${\rm c}$, being indispensable for achieving high recall, receives a higher contribution score from the LC approach to reflect its pivotal role in ensuring a satisfactory performance floor.

Conversely, SV, grounded in the principle of average marginal contribution, offers a different perspective. It evaluates each participant's contribution by averaging their marginal impact across all possible coalition formations. While Participant ${\rm c}$'s marginal contributions are undoubtedly significant, SV also considers the positive, albeit smaller and less critical, marginal contributions of Participant ${\rm a}$ and ${\rm b}$ in various coalitions. By averaging these marginal contributions across all permutations, SV provides a more `moderate' or `balanced' evaluation of contribution. Although Participant ${\rm c}$ remains highly valued due to its substantial impact, the contributions of ${\rm a}$ and ${\rm b}$ are also acknowledged and factored into the final evaluation, leading to a relatively lower contribution score for ${\rm c}$ compared to the LC, and a relatively higher score for ${\rm a}$.

To validate the robustness of the above finding, we conducted multiple experimental runs with varying random seeds. The distributions of the LC and the SV contributions for each participant across these runs are visualized as box plots in Fig. \ref{LV_SC}. These box plots robustly demonstrate that the LC consistently allocate a higher contribution to Participant ${\rm c}$ and a lower contribution to Participant ${\rm a}$ compared to the SV.

The allocation pattern provided by the LC approach \textbf{appears to be more aligned with intuitive expectations} when considering the nature of their respective features. Recall that Participant ${\rm c}$ represents a specialized cardiology department, possessing the features more relevant to heart disease risk. In contrast, Participant ${\rm a}$, holding only basic demographic features like `age' and `sex', contributes less directly to the core task of heart disease diagnosis. Therefore, the contribution evaluation provided by the LC suggests a potentially more suitable approach for FL networks, particularly in applications like VFL where feature importance is inherently unevenly distributed.

\section{Conclusions}

In this paper, we addressed the critical challenge of designing a fair, stable, and scalable payoff allocation mechanism for FL ecosystems. We proposed a framework based on the LC concept, which prioritizes the stability of the grand coalition by minimizing the dissatisfaction of the most vulnerable subgroups. To overcome the exponential computational complexity inherent in cooperative game theory, we developed a practical stack-based pruning algorithm that drastically reduces the number of required model evaluations without compromising allocation quality.

Our extensive case studies in intrusion detection and smart healthcare demonstrate the superiority of the LC approach. It outperforms traditional metrics like data volume and leave-one-out by recognizing strategic bargaining power and preventing coalition defection. Furthermore, distinct from the Shapley Value which averages marginal contributions, the LC mechanism places greater emphasis on participant indispensability. This characteristic offers a more intuitive and robust valuation, particularly in vertical FL scenarios where feature importance varies significantly. Ultimately, this work provides a computationally feasible path toward sustainable, large-scale cross-silo FL, ensuring that all participants remain incentivized to collaborate in dynamic industrial environments.



\bibliographystyle{IEEEtran}
\bibliography{IEEEabrv,mybibfile}

\end{document}